\documentclass{article}
\usepackage{amssymb}

\usepackage{graphicx}
\usepackage{amsmath}

\input{tcilatex}

\begin{document}

\title{\textsf{Spin} \textsf{coherent states with monopole harmonics on the Riemann
sphere for the Kravchuk oscillator }}
\author{Zouha\"{i}r Mouayn}
\date{{\small PO.Box 123, B\'{e}ni Mellal, Morocco }\\
{\small mouayn@gmail.com}}
\maketitle

\begin{abstract}
We consider a class of generalized spin coherent states by choosing the
labeling coefficients to be monopole harmonics. The latters are $L^{2}$
eigenstates of the $m$th spherical Landau level on the Riemann sphere with $%
m\in \mathbb{Z}_{+}$. We verify that the Klauder's minimum properties for
these states to be considered as coherent states are satisfied. We
particularize them for the case of the Kravchuk oscillator and we obtain
explicite expression for their wave functions. The associated coherent
states transforms provide us with a Bargmann-type representation for the
states of the oscillator Hilbert space. For the lowest level $m=0$ indexing
monopole harmonics$,$ we identify the obtained coherent states to be those
of \ Klauder-Perelomov type which were constructed in Ref. $\left[
J.Math.Phys.48,112106\left( 2007\right) \right] .$
\end{abstract}

\section{Introduction}

The spin $SU\left( 2\right) $ coherent states $\left( SCS\right) $ were
introduced in the early 1970's by Radcliffe $\left[ 1\right] ,$ Gilmore $%
\left[ 2\right] $ and Perelomov $\left[ 3\right] .$ They are also named 
\textit{atomic }or\textit{\ Bloch} coherent states. This diversity of
appellations reflects the range of domains in quantum physics where these
objects play some role. One can introduce these states by following a
probabilistic and Hilbertian scheme as explained in details in\ the book of
Gazeau $\left[ 4\right] .$ Precisely, the SCS are the field states that are
superposition of the number states with appropriately chosen coefficients.
These labelling coefficients are such that the associated photon-counting
distribution is a binomial probability distribution $\left[ 5\right] .$ \ In
addition, these coefficients constitute an orthonormal basis of a Hilbert
space of analytic functions on the Riemann sphere satisfying a certain
growth condition.

Now, as in $\left[ 6\right] ,$ we replace the usual labelling coefficients
in the superposition defining the SCS by an orthonormal basis consisting of
monopole harmonics that are $L^{2}$ eigenfunctions of an invariant Laplacian
on the Riemann sphere corresponding to discrete eigenvalues (\textit{%
spherical Landau levels}) to introduce a class of generalized spin coherent
states (GSCSs)$.$ Each of these eigenvalue is of\textit{\ finite}
degeneracy. Here, we precisely verify that the basic minimum properties for
the constructed states to be considered as coherent states are satisfied.
Namely, the conditions which have been formulated by Klauder $\left[ 7\right]
$ $:\left( a\right) $ the continuity of labelling, $\left( b\right) $ the
fact that these states are normalizable but not orthogonal and $\left(
c\right) $ these states fulfilled the resolution of the identity with a
positive weight function. \ Next, we particularize the GSCSs formalism for
the case of the Hamiltonian of the Kravchuk oscillator $\left[ 8\right] $
which is a\textit{\ finite} model oscillator whose importance consists in
the fact that it can be considered as a discrete analogue of the harmonic
oscillator and that its eigenfunctions which are the kravchuk functions
coincide with the harmonic oscillator functions in a certain limit. Next, we
obtain explicitly the wave functions of these GSCS which enable us to write
a Bargmann-type representation of any state in the oscillator Hilbert space.
For the lowest level $m=0$ indexing monopole harmonics$,$ we identify the
obtained coherent states to be those of \ Klauder-Perelomov type which were
constructed in $\left[ 9\right] .$

The paper is organized as follows. Section 2 deals with some needed facts on
the monopole harmonics. In Section 3, we discuss a class of generalized spin
coherent states attached to monopole harmonics and verify that they satisfy
the basic minimum properties of coherent states. In Section 4, we recall
briefly the definition of the Kravchuk oscillator and its eigenstates. In
section 5, we particularize the constructed coherent states for the case of
the Kravchuk oscillator and we discuss a Bargmann-type representation for
the states of the oscillator Hilbert space. In section 6, we focus on a
particular case of the constructed coherent states, for which we establish a
connection with known results.

\section{Monopole harmonics on the Riemann sphere}

We shall recall here some needed spectral properties the Dirac monopole
Hamiltonian operator according to references $\left[ 10-11\right] .$ For
this, we start by \ identifying the sphere $\mathbb{S}^{2}$ with the
extended complex plane $\mathbb{C}\mathbf{\cup }\left\{ \infty \right\}
\equiv \overline{\mathbb{C}}$, called the Riemann sphere, via the
stereographic coordinate $z=x+iy$, $x,y\in \mathbb{R}.$ We shall work within
a fixed coordinate neighborhood with coordinate $z$\ obtained by deleting
the ''point at infinity'' $\left\{ \infty \right\} .$\ Near this point we
use instead of $z$\ the coordinate $z^{-1}.$\ \ In the stereographic
coordinate $z$, the Hamiltonian operator of the Dirac monopole with charge $%
q=2\nu $ in the notations of Veslov and Ferepontov $(\left[ 10\right] ,$ $%
p.598)$ reads 
\begin{equation}
H_{2\nu }=-\left( 1+z\overline{z}\right) ^{2}\frac{\partial ^{2}}{\partial
z\partial \overline{z}}-\nu z\left( 1+z\overline{z}\right) \frac{\partial }{%
\partial z}+\nu \overline{z}\left( 1+z\overline{z}\right) \frac{\partial }{%
\partial \overline{z}}+\nu ^{2}\left( 1+z\overline{z}\right) -\nu ^{2}. 
\tag{2.1}
\end{equation}
This operator acts on the sections of the $U\left( 1\right) -$bundle with
the first Chern class $q.$ We have denoted by $\nu \geq 0$\ the strength of
the quantized magnetic field. The associated annihilation and creation
operators are given by 
\begin{eqnarray}
A_{\nu } &:&=\left( 1+\left| z\right| ^{2}\right) \frac{\partial }{\partial 
\overline{z}}+\nu z;  \TCItag{2.2} \\
A_{\nu }^{\ast } &:&=-\left( 1+\left| z\right| ^{2}\right) \frac{\partial }{%
\partial z}+\left( \nu +1\right) \overline{z}  \TCItag{2.3}
\end{eqnarray}
The Hamiltonian $H_{2\nu }$ in (2.1) is acting in the Hilbert space $%
L^{2}\left( \mathbb{S}^{2},d\mu \right) $\textbf{\ }with $d\mu \left(
z\right) =\left( 1+z\overline{z}\right) ^{-2}d\eta \left( z\right) ,$ $d\eta
\left( z\right) =\pi ^{-1}dxdy$ being the Lebesgue measure on $\mathbb{C}%
\mathbf{.}$ \ To find the ground state, the authors in $\left[ 10\right] $
solved the equation $A_{\nu }\left[ \psi \right] =0$ and found that $\psi $
should be of the form 
\begin{equation}
\psi \left( z\right) =\left( 1+z\overline{z}\right) ^{-\nu }\phi \left(
z\right)  \tag{2.4}
\end{equation}
where $\phi \left( z\right) $ is any holomophic function of $z.$ Because of
the condition $\psi \in L^{2}\left( \mathbb{S}^{2}\right) ,$ the function $%
\phi \left( z\right) $ must be a polynomial of order $\leq 2\nu .$ This
gives a space of dimension $2\nu +1.$ Next, by making use of the operator $%
A_{\nu }^{\ast }$ they have obtained that the eigenfunctions of the Dirac
monopole $H_{2\nu }$ corresponding to the eigenvalue 
\begin{equation}
\lambda _{\nu ,m}:=\left( 2m+1\right) \nu +m\left( m+1\right) \text{, }%
m=0,1,2,...  \tag{2.5}
\end{equation}
form a space of dimension $2\nu +2m+1$ and are described as 
\begin{equation}
\Phi \left( z\right) =A_{\nu -m}^{\ast }...A_{\nu -2}^{\ast }A_{\nu
-1}^{\ast }\left( 1+z\overline{z}\right) ^{-\nu -m}P\left( z\right) . 
\tag{2.6}
\end{equation}
and $P\left( z\right) $ is a polynomial of degree $\leq 2\nu +2m.$ The
eigenfunctions of the Dirac monopole are known as monopole harmonics and
have been investigated by Wu and Yang $\left[ 12\right] ,$ who were probably
the first to identify them as sections. \ To obtain explicit expressions of
these eigen\textit{sections} in the coordinate $z,$ one also can make use of
the results obtained by Peetre and Zhang $\left[ 13\right] $ first by
establishing an intertwining relation between the shifted operator 
\begin{equation}
\mathsf{\ }H_{2\nu }-\nu =\left( 1+z\overline{z}\right) ^{-\nu }\Delta
_{2\nu }\left( 1+z\overline{z}\right) ^{\nu },  \tag{2.7}
\end{equation}
and an invariant Laplacian 
\begin{equation}
\Delta _{2\nu }:=-\left( 1+z\overline{z}\right) ^{2}\frac{\partial ^{2}}{%
\partial z\partial \overline{z}}+2\nu \overline{z}\left( 1+z\overline{z}%
\right) \frac{\partial }{\partial \overline{z}}  \tag{2.8}
\end{equation}
acting in the Hilbert space $L^{2,\nu }\left( \mathbb{S}^{2}\right)
:=L^{2}\left( \mathbb{S}^{2},d\mu _{\nu }\left( z\right) \right) $ with $%
d\mu _{\nu }\left( z\right) =\left( 1+z\overline{z}\right) ^{-2-2\nu }d\nu
\left( z\right) .$ \ According to (2.7) any ket $\mid \phi >$ of $\ L^{2,\nu
}\left( \mathbb{S}^{2}\right) $ is represented by

\begin{equation}
\left( 1+z\overline{z}\right) ^{-\nu }<z\mid \phi >\text{in }L^{2}\left( 
\mathbb{S}^{2}\right) .  \tag{2.9}
\end{equation}
Next by ($\left[ 13\right] ,$ $pp.228-229$), the relations (2.7), (2.9) and
with the help of the polynomial functions 
\begin{equation}
\text{ }Q_{j}^{\nu ,m}\left( u\right) =\left( j!\right) ^{-1}\left(
m+j\right) !\text{ }_{2}F_{1}\left( -m,2\nu +m+1,j+1;u\right)  \tag{2.10}
\end{equation}
where $_{2}F_{1}$ is the Gauss hypergeometric function $\left[ 14\right] $,
one can check that the functions

\begin{equation}
\Phi _{j}^{\nu ,m}\left( z\right) :=\left( 1+z\overline{z}\right) ^{-\nu
}z^{j}\text{ }Q_{j}^{\nu ,m}\left( \frac{z\overline{z}}{1+z\overline{z}}%
\right) ,\text{ \quad }-m\leq j\leq 2\nu +m  \tag{2.11}
\end{equation}
constitutes an orthogonal set in the eigenspace 
\begin{equation}
\mathcal{A}_{m}^{\nu }\left( \mathbb{S}^{2}\right) =\left\{ \Phi \in
L^{2}\left( \mathbb{S}^{2}\right) ,H_{2\nu \text{ }}\Phi =\lambda _{\nu ,m}\
\Phi \right\}  \tag{2.12}
\end{equation}
corresponding to the eigenvalue $\lambda _{\nu ,m}$\ given in (2.5). But
since the terminating $_{2}F_{1}$-sum can be written as 
\begin{equation}
_{2}F_{1}\left( k+\beta +\alpha +1,-k,1+\alpha ;\frac{1-t}{2}\right) =\frac{%
k!\Gamma \left( 1+\alpha \right) }{\Gamma \left( k+1+\alpha \right) }%
P_{k}^{(\alpha ,\beta )}\left( t\right)  \tag{2.13}
\end{equation}
in terms of the Jacobi polynomial ($\left[ 14\right] ):$ 
\begin{equation}
P_{k}^{(\alpha ,\beta )}\left( t\right) =\sum\limits_{s=0}^{k}\left( 
\begin{array}{c}
k+\alpha \\ 
k-s
\end{array}
\right) \left( 
\begin{array}{c}
k+\beta \\ 
s
\end{array}
\right) \left( \frac{t-1}{2}\right) ^{s}\left( \frac{t+1}{2}\right) ^{k-s} 
\tag{2.14}
\end{equation}
then one can also present an orthonormal basis of the space $\left(
2.12\right) $ by the expression 
\begin{equation}
\widetilde{\Phi }_{j}^{\nu ,m}\left( z\right) :=\gamma _{j}^{\nu ,m}\left(
1+z\overline{z}\right) ^{-\nu }z^{j}P_{m}^{\left( j,2\nu -j\right) }\left( 
\frac{1-z\overline{z}}{1+z\overline{z}}\right)  \tag{2.15}
\end{equation}
where $-m\leq j\leq 2\nu +m$ and the constant is given by 
\begin{equation}
\gamma _{j}^{\nu ,m}:=\sqrt{\frac{\left( 2\nu +2m+1\right) \left( 2\nu
+m\right) !m!}{(m+j)!(2\nu +m-j)!}}.  \tag{2.16}
\end{equation}
We end this section by the following remarks.

\textbf{Remark 3.1. }In $\left[ 10\right] ,$ the authors have also pointed
out that if $2\nu $ is even integer then $H_{2\nu }$ can be intertwined with
the standard Laplace-Beltrami operator $-\Delta _{\mathbb{S}^{2}}$ on the
sphere as follows: 
\begin{equation}
H_{2\nu }D=D\left( -\Delta _{\mathbb{S}^{2}}-\nu ^{2}\right) ,D=D_{\nu
-1}D_{\nu -2}...D_{1}D_{0},  \tag{2.18}
\end{equation}
where $D_{\nu }$:$=\left( 1+z\overline{z}\right) \overline{\partial }+\nu z.$
So that one can use the eigenfunctions of $-\Delta _{\mathbb{S}^{2}},$ which
are spherical harmonics, to construct the eigenfunctions of $H_{2\nu }.$

\textbf{Remark 3.2. }Note that\textbf{\ }the strength of the magnetic field $%
\nu $ should satisfy $\nu \in \left\{ \frac{1}{2},1,\frac{3}{2},...\right\} $%
. As matter of fact, there is a result \ called Dirac's quantization for
monopole charges which requires that the total flux of the magnetic field
across a closed surface must be quantized, i.e. it must be an integer
multiple of a universal constant. This result is about cohomology groups for
hermitian line bundles $\left[ 15\right] $ and is also known as the
Weil-Souriau-Kostant quantization condition $\left[ 16\right] .$ For more
information on Dirac monopoles, see $\left[ 17\right] .$

\section{Generalized spin coherent states with monopole harmonics}

Here, we shall make use of the same notation $\nu $ to consider a Hilbert
space denoted $\mathcal{H}$ of dimension $\left( 2\nu +1\right) ,$ carrying
an irreducible representation of the group $SU\left( 2\right) .$ Each space $%
\mathcal{H}$ is associated with a spin of length $\nu \in \left\{ \frac{1}{2}%
,1,\frac{3}{2},...\right\} .$ To introduce spin-coherent states (SCS), it is
convenient to select states of highest (lowest) weight $\mid \pm \nu >$ as
reference states$.$ These states are invariant under a change of phase,
hence the isotopy group is given by $U\left( 1\right) $. Therefore the
closed space $SU\left( 2\right) /U\left( 1\right) $ is the surface of the
sphere $\mathbb{S}^{2}$ identified with $\overline{\mathbb{C}}$ as mentioned
above, which correspond to the phase space of a classical spin. The SCS
defined in this Hilbert space are given by ($\left[ 1\right] ,$p.315)$:$%
\begin{equation}
\mid z,\nu ,0>=\left( 1+z\overline{z}\right) ^{-\nu }\sum_{k=0}^{2\nu }\sqrt{%
\frac{\Gamma \left( 2\nu +1\right) }{k!\Gamma \left( 2\nu -k+1\right) }}%
\left| z\right| ^{k}e^{ik\theta }\mid k,\nu >  \tag{3.1}
\end{equation}
where the labelling parameter $z=\left| z\right| e^{i\theta }\in \overline{%
\mathbb{C}}$ and $\mid k,\nu >$ are number states of the field mode. Due to
the fact that the probability for the production of $k$ photons, given by
the quantity 
\begin{equation}
\left| <k,\nu \mid z,\nu ,0>\right| ^{2}=\frac{\Gamma \left( 2\nu +1\right) 
}{k!\Gamma \left( 2\nu -k+1\right) }\left( z\overline{z}\right) ^{k}\left(
1+z\overline{z}\right) ^{-2\nu }  \tag{3.2}
\end{equation}
is a binomial probability density $\frak{B}\left( n,\tau \right) $ with $%
n:=2\nu $ as an integer parameter and $\tau :=z\overline{z}\left( 1+z%
\overline{z}\right) ^{-1}$ as a Bernoulli parameter with $0<\tau <1$, the
SCS in $\left( 3.1\right) $ is exactly of the form of the binomial state $%
\left[ 18\right] .$ These states in $\left( 3.1\right) $ constitute a
resolution of the identity of the Hilbert space

\begin{equation}
\int\limits_{\mathbb{C}}\mid z,\nu ,0><0,\nu ,z\mid \frac{\left( 2\nu
+1\right) }{\left( 1+z\overline{z}\right) ^{2}}=\mathbf{1}_{\mathcal{H}} 
\tag{3.3}
\end{equation}
As pointed out in ($\left[ 4\right] $, p.$81$) one should notice here the
similarity with the standard coherent states 
\begin{equation}
\mathbb{C\ni }\zeta \mapsto \mid \zeta >=\sum_{n=0}^{+\infty }\exp \left( -%
\frac{1}{2}\zeta \overline{\zeta }\right) \frac{\zeta ^{n}}{\sqrt{n!}}\mid n>
\tag{3.4}
\end{equation}
which are obtained from the spin CS at the limit of high spin $N=2\nu $
through a contraction process. The latter is carried out through a scaling
of the complex variable $z,$ namely $\zeta =\sqrt{N}z$ and $n=2\nu -k,$ $%
\mid k,\nu >\equiv \mid n>$ and the limit $N\rightarrow \infty :$%
\begin{equation}
\mid z=\frac{\zeta }{\sqrt{N}}>_{spin}\rightarrow \mid \zeta >.  \tag{3.5}
\end{equation}
We also note that the coefficients in $\left( 3.1\right) $ can be written as 
\begin{equation}
z\mapsto \sqrt{\frac{\Gamma \left( 2\nu +1\right) }{j!\Gamma \left( 2\nu
-j+1\right) }}\left( 1+z\overline{z}\right) ^{-\nu }z^{j}=\left( 2\nu
+1\right) ^{-\frac{1}{2}}\widetilde{\Phi }_{j}^{\nu ,0}\left( z\right) 
\tag{3.6}
\end{equation}
in terms of the elements $\widetilde{\Phi }_{j}^{\nu ,0}\left( z\right) $ of
the orthonormal basis of the space $\mathcal{A}_{0}^{\nu }\left( \mathbb{S}%
^{2}\right) =\left\{ \Phi \in L^{2}\left( \mathbb{S}^{2}\right) ,H_{2s\text{ 
}}\Phi =\lambda _{\nu ,0}\Phi \right\} $ in $\left( 2.1\right) $
corresponding to the lowest energy level $\lambda _{\nu ,0}.$ The latter is
obtained by setting for $m=0$ in Eq.$\left( 2.5\right) .$ From this
observation, we propose as in $\left[ 6\right] $ a generalization for these
spin coherent states by following the probabilistic and Hilbertian scheme as
explained in $\left( \left[ 4\right] \text{, p.74}\right) .$ More precisely,
we state the following.

\textbf{Definition 3.1. }\textit{For each fixed }$m\in \mathbb{Z}_{+}$ 
\textit{and} $\nu \in \left\{ \frac{1}{2},1,\frac{3}{2},...\right\} $\textit{%
\ a class of generalized spin coherent state (GSCS}$)$\textit{\ is defined\
by the form}

\begin{equation}
\mid z,\nu ,m>:=\left( \mathcal{N}_{\nu ,m}\left( z\right) \right) ^{-\frac{1%
}{2}}\sum_{j=-m}^{2\nu +m}\overline{\widetilde{\Phi }_{j}^{\nu ,m}\left(
z\right) }\mid j,\nu >  \tag{3.7}
\end{equation}
\textit{where }$\mathcal{N}_{\nu ,m}\left( z\right) $\textit{\ is a
normalization factor and }$\widetilde{\Phi }_{j}^{\nu ,m}\left( .\right) $%
\textit{\ is the monopole harmonic defined in (2.15). }

Now, one of the important task to achieve is to determine explicitly the
overlap relation between two GSCSs.

\textbf{Proposition 3.1.} \textit{Let} $m\in \mathbb{Z}_{+}$ \textit{and} $%
\nu \in \left\{ \frac{1}{2},1,\frac{3}{2},...\right\} .$ \textit{Then, for
every} $z,w\in \overline{\mathbb{C}},$ \textit{the overlap relation between
two GSCSs is given through the scalar product}

\begin{equation}
<w,\nu ,m\mid z\text{ },\nu ,m>_{\mathcal{H}}=\frac{(\nu +2m+1)(1+z\overline{%
w})^{2\nu }}{\left( \mathcal{N}_{\nu ,m}\left( z\right) \mathcal{N}_{\nu
,m}\left( w\right) \right) ^{\frac{1}{2}}\left( 1+z\overline{z}\right) ^{\nu
}\left( 1+w\overline{w}\right) ^{\nu }}.  \tag{3.8}
\end{equation}
\begin{equation*}
\times _{2}\digamma _{1}\left( -m,m+2\nu +1,1;\frac{\left( z-w\right) \left( 
\overline{z}-\overline{w}\right) }{(1+z\overline{z})\left( 1+w\overline{w}%
\right) }\right)
\end{equation*}
\textit{where }$_{2}\digamma _{1}$\textit{\ is a terminating Gauss
hypergeometric sum}.

\textbf{Proof. }In view of Eq.$\left( 3.7\right) $, the scalar product of
two GSCS $\mid z$ $,\nu ,m>$ and $\mid w$ $,\nu ,m>$ in $\mathcal{H}$ reads 
\begin{equation}
<w,\nu ,m\mid z\text{ },\nu ,m>_{\mathcal{H}}=\left( \mathcal{N}_{\nu
,m}\left( z\right) \mathcal{N}_{\nu ,m}\left( w\right) \right) ^{-\frac{1}{2}%
}\frak{S}_{z,w}^{\nu ,m}  \tag{3.9}
\end{equation}
where 
\begin{equation}
\frak{S}_{z,w}^{\nu ,m}=\sum\limits_{j=-m}^{2\nu +m}\widetilde{\Phi }%
_{j}^{\nu ,m}\left( z\right) \overline{\widetilde{\Phi }_{j}^{\nu ,m}\left(
w\right) }.  \tag{3.10}
\end{equation}
Recalling Eq.$\left( 2.11\right) $, we can rewrite the finite sum $\left(
3.10\right) $ in terms of the product of the polynomial functions $%
Q_{j}^{\nu ,m}\left( u\right) $ in $\left( 2.10\right) $. Next, we use the
addition formula due to J. Peetre and G. Zhang $\left( \left[ 13\right] ,%
\text{p.231}\right) $ involving the functions in $\left( 2.10\right) $ to
obtain that 
\begin{equation}
\frak{S}_{z,w}^{\nu ,m}=\frac{\left( 2\nu +2m+1\right) \left( 1+z\overline{w}%
\right) ^{2\nu }}{\left( \left( 1+z\overline{z}\right) \left( 1+w\overline{w}%
\right) \right) ^{\nu }}._{2}\digamma _{1}\left( -m,m+2\nu +1,1;\frac{\left(
z-w\right) \left( \overline{z}-\overline{w}\right) }{\left( 1+z\overline{z}%
\right) \left( 1+w\overline{w}\right) }\right)  \tag{3.11}
\end{equation}
Returning back to Eq.$\left( 3.9\right) $ and inserting the expression $%
\left( 3.11\right) $ we arrive at the announced result.$\blacksquare $

\textbf{Corollary 3.1}. \textit{The normalization factor in}$\mathit{\ }%
\left( 3.7\right) $\textit{\ is given by} 
\begin{equation}
\mathcal{N}_{\nu ,m}\left( z\right) =2\left( \nu +m\right) +1  \tag{3.12}
\end{equation}
\textit{for every} $z\in \overline{\mathbb{C}}.$

\textbf{Proof.\ }We first make appeal to the relation $\left( 2.13\right) $
connecting the $_{2}\digamma _{1}-$sum with the Jacobi polynomial to rewrite
Eq.$\left( 3.8\right) $ as 
\begin{eqnarray}
\sqrt{\mathcal{N}_{\nu ,m}\left( z\right) \mathcal{N}_{\nu ,m}\left(
w\right) } &=&\frac{\left( 1+z\overline{z}\right) ^{-\nu }(2\nu +2m+1)(1+z%
\overline{w})^{2\nu }}{\left( 1+w\overline{w}\right) ^{\nu }<w,\nu ,m\mid z%
\text{ },\nu ,m>_{\mathcal{H}}}  \TCItag{3.13} \\
&&\times P_{m}^{\left( 0,2\nu \right) }\left( 1-\frac{2\left( z-w\right)
\left( \overline{z}-\overline{w}\right) }{(1+z\overline{z})\left( 1+%
\overline{w}w\right) }\right) .  \notag
\end{eqnarray}
The factor $\mathcal{N}_{\nu ,m}\left( z\right) $ should be such that $%
<z,\nu ,m\mid z,\nu ,m>_{\mathcal{H}}=1.$ So that we put $z=w$ in $\left(
3.13\right) $ to obtain the expression 
\begin{equation}
\mathcal{N}_{\nu ,m}\left( z\right) =\left( 2\nu +2m+1\right) P_{m}^{\left(
0,2\nu \right) }\left( 1\right) .  \tag{3.14}
\end{equation}
Finally, we apply the fact that $\left( \left[ 14\right] \text{, p.57}%
\right) :$%
\begin{equation}
P_{n}^{\left( \alpha ,\sigma \right) }\left( 1\right) =\frac{\Gamma \left(
n+\alpha +1\right) }{n!\Gamma \left( \alpha +1\right) }  \tag{3.15}
\end{equation}
in the case of $\alpha =0,n=m$ and $\sigma =2\nu .$ This ends the proof.$%
\blacksquare $

\textit{\ }

\textbf{Proposition 3.2.}\textit{\ Let}\textbf{\ }$m\in \mathbb{Z}_{+}$ 
\textit{and }$\nu \in \left\{ \frac{1}{2},1,\frac{3}{2},...\right\} $. 
\textit{Then,} \textit{the GSCS }$\mid z,\nu ,m>$\textit{\ satisfy the
following resolution of the identity} 
\begin{equation}
\int\limits_{\mathbb{C}}\mid z,\nu ,m><z,\nu ,m\mid d\mu _{\nu ,m}\left(
z\right) =\mathbf{1}_{\mathcal{H}}  \tag{3.16}
\end{equation}
\textit{where}\textbf{\ }$\mathbf{1}_{_{\mathcal{H}}}$\textit{\ is the
identity operator and }$d\mu _{\nu ,m}\left( z\right) $\textit{\ is a
measure which can be expressed through a Meijer's }$G$ \textit{function as} 
\textit{\ } 
\begin{equation}
d\mu _{\nu ,m}\left( z\right) :=\left( 2\nu +2m+1\right) G_{11}^{11}\left( z%
\overline{z}\mid 
\begin{array}{c}
-1 \\ 
0
\end{array}
\right) d\eta \left( z\right) ,  \tag{3.17}
\end{equation}
\textit{where} $d\eta \left( z\right) $ \textit{denotes the Lebesgue measure
on }$\mathbb{C}$.

\textbf{Proof}. We assume that the measure takes the form $d\mu _{\nu
,m}\left( z\right) =\mathcal{N}_{\nu ,m}\left( z\right) \Omega \left(
z\right) d\eta \left( z\right) $ where $\Omega \left( z\right) $ is an
auxiliary density to be determined. Let $\varphi \in \mathcal{H}$ and let us
start by writing the following action 
\begin{equation}
\mathcal{O}\left[ \varphi \right] :=\left( \int\limits_{\mathbb{C}}\mid
z,\nu ,m><z,\nu ,m\mid d\mu _{\nu ,m}\left( z\right) \right) \left[ \varphi %
\right]  \tag{3.18}
\end{equation}
\begin{equation}
=\int\limits_{\mathbb{C}}<\varphi \mid z,\nu ,m><z,\nu ,m\mid d\mu _{\nu
,m}\left( z\right)  \tag{3.19}
\end{equation}
Making use of Eq.$\left( 3.7\right) ,$ we obtain successively 
\begin{equation}
\mathcal{O}\left[ \varphi \right] =\int\limits_{\mathbb{C}}<\varphi \mid
\left( \mathcal{N}_{\nu ,m}\left( z\right) \right) ^{-\frac{1}{2}%
}\sum_{j=-m}^{2\nu +m}\overline{\widetilde{\Phi }_{j}^{\nu ,m}\left(
z\right) }\mid j,\nu >><z,\nu ,m\mid d\mu _{\nu ,m}\left( z\right) 
\tag{3.20}
\end{equation}
\begin{equation}
=\left( \sum_{j,k=-m}^{2\nu +m}\int\limits_{\mathbb{C}}\overline{\widetilde{%
\Phi }_{j}^{\nu ,m}\left( z\right) }\widetilde{\Phi }_{k}^{\nu ,m}\left(
z\right) \mid j,\nu ><k,\nu \mid \left( \mathcal{N}_{\nu ,m}\left( z\right)
\right) ^{-1}d\mu _{\nu ,m}\left( z\right) \right) \left[ \varphi \right] . 
\tag{3.21}
\end{equation}
Replace $d\mu _{\nu ,m}\left( z\right) $ by $\mathcal{N}_{\nu ,m}\left(
z\right) \Omega \left( z\right) d\nu \left( z\right) ,$ then Eq.$\left(
3.21\right) $ takes the form 
\begin{equation}
\mathcal{O}=\sum_{j,k=-m}^{2\nu +m}\left[ \int\limits_{\mathbb{C}}\overline{%
\widetilde{\Phi }_{j}^{\nu ,m}\left( z\right) }\widetilde{\Phi }_{k}^{\nu
,m}\left( z\right) \Omega \left( z\right) d\nu \left( z\right) \right] \mid
j,\nu ><k,\nu \mid .  \tag{3.22}
\end{equation}
Then, we need to obtain 
\begin{equation}
\int\limits_{\mathbb{C}}\overline{\widetilde{\Phi }_{j}^{\nu ,m}\left(
z\right) }\widetilde{\Phi }_{k}^{\nu ,m}\left( z\right) \Omega \left(
z\right) d\nu \left( z\right) =\delta _{j,k}.  \tag{3.23}
\end{equation}
For this we recall the orthogonality relation satisfied by monopole
harmonics in $\left( 2.15\right) $ in $L^{2}\left( \overline{\mathbb{C}}%
,\left( 1+z\overline{z}\right) ^{-2}d\eta \left( z\right) \right) $\textbf{, 
} as 
\begin{equation}
\int\limits_{\overline{\mathbb{C}}}\overline{\widetilde{\Phi }_{j}^{\nu
,m}\left( z\right) }\widetilde{\Phi }_{k}^{\nu ,m}\left( z\right) \left( 1+z%
\overline{z}\right) ^{-2}d\eta \left( z\right) =\delta _{j,k}.  \tag{3.24}
\end{equation}
This suggests us to set $\Omega \left( z\right) :=\left( 1+z\overline{z}%
\right) ^{-2}d\eta \left( z\right) .$ By making us of the identity $\left( %
\left[ 19\right] \right) :$ 
\begin{equation}
G_{11}^{11}\left( u\mid 
\begin{array}{c}
a \\ 
b
\end{array}
\right) =\Gamma \left( 1-a+b\right) u^{b}\left( 1+u\right) ^{a-b-1} 
\tag{3.25}
\end{equation}
for $u=z\overline{z}$, $a=-1$ and $b=0$, we arrive at the announced form for
the measure $d\mu _{\nu ,m}$ in $\left( 3.17\right) .$ Because of this this
measure, Eq.$\left( 3.22\right) $ reduces to 
\begin{equation}
\mathcal{O}=\sum\limits_{j,k=-m}^{2\nu +m}\delta _{j,k}\mid j,\nu ><k,\nu
\mid =\mathbf{1}_{\mathcal{H}}.  \tag{3.26}
\end{equation}
This ends the proof.$\blacksquare $

\textbf{Remark 3.1.} Note that when $m=0$, Eq.$\left( 3.16\right) $ leads to
Eq.$\left( 3.3\right) .$ For $m\neq 0$ , the fact that we have written the
measure $d\mu _{\nu ,m}\left( z\right) $ in $\left( 3.17\right) $ in terms
of the Meijer's G-function could be of help when tackling the ''\textit{%
photon-added coherent states} (\textit{PACS})'' problem for the GSCS under
consideration.

\textbf{Proposition 3.3. }\textit{The states }$\mid z,\nu ,m>$\textit{\
satisfy the continuity property with respect to the labelling point }$z.$%
\textit{\ That is, the norm of the difference of two states } 
\begin{equation}
\rho _{\nu ,m}\left( z,w\right) \mathit{\ }:=\left\| \mid z,\nu ,m>-\mid
w,\nu ,m>\right\| _{\mathcal{H}}  \tag{3.28}
\end{equation}
\textit{goes} \textit{to zero} \textit{whenever }$z\rightarrow w.$

\textbf{Proof. }By using the fact that any GSCS is normalized by the factor
given in $\left( 3.12\right) $, a direct calculation enables us to write the
square of the quantity in $\left( 3.28\right) $ as 
\begin{equation}
\left( \mathit{\ }\rho _{\nu ,m}\left( z,w\right) \mathit{\ }\right)
^{2}=2\left( 1-\func{Re}<z,\nu ,m\mid w,\nu ,m>\right) .  \tag{3.29}
\end{equation}
So it is clear that when $z\rightarrow w,$ the terminating Gauss
hypergeometric function goes to $1$ and the prefactor in $\left( 3.8\right) $
goes to $\left( 2\nu +2m+1\right) .$ Therefore,the overlap takes the value $%
1 $ and consequently $\rho _{\nu ,m}\left( z,w\right) \rightarrow 0$ in $%
\left( 3.29\right) .\blacksquare $

As we can see, these GSCS are independent of the basis $\mid j,\nu >$ we use
and the only condition which is implicitly fulfilled is the orthonormality
of the basis vectors of $\mathcal{H}$. But if we want to attach our GSCS to
a concrete quantum system then a Hamiltonian operator should be specified
together with a corresponding explicit eigenstates basis. This will be the
goal of the next section.

\section{The Kravchuk oscillator}

The Kravchuk polynomials $K_{k}^{\left( p\right) }\left( x,N\right) $ of
degree $k=0,1,2,...,N,$ in the variable $x\in \left[ 0,N\right] $ and of the
parameter $0<p<1,$ are related to the binomial probability distribution $%
\left[ 5\right] .$ They satisfy the well known three-term recurrence
relation $:$ 
\begin{equation}
\left( x-k-p\left( N-2k\right) \right) K_{k}^{\left( p\right) }\left(
x,N\right)  \tag{4.1}
\end{equation}
\begin{equation*}
=\left( k+1\right) K_{k+1}^{\left( p\right) }\left( x,N\right) +p\left(
1-p\right) \left( N-k+1\right) K_{k-1}^{\left( p\right) }\left( x,N\right)
\end{equation*}
and can be defined in terms of the Gauss hypergeometric function through 
\begin{equation}
K_{k}^{\left( p\right) }\left( x,N\right) :=\left( -1\right) ^{k}p^{k}\left( 
\begin{array}{c}
N \\ 
k
\end{array}
\right) ._{2}\digamma _{1}\left( -k,-x,-N;\frac{1}{p}\right) .  \tag{4.2}
\end{equation}
For each fixed nonzero positive integer $N,$ the $N+1$ Kravchuk polynomials $%
\left\{ K_{k}^{\left( p\right) }\left( x,N\right) \right\} _{k=0}^{N}$ are
an orthogonal set \ with \ respect to a discrete weight function with finite
support, namely 
\begin{equation}
\sum_{j=0}^{N}\varrho \left( j\right) K_{k}^{\left( p\right) }\left(
j,N\right) K_{n}^{\left( p\right) }\left( j,N\right) =\frac{N!}{k!\left(
N-k\right) !}\left( p\left( 1-p\right) \right) ^{n}\delta _{k,n}  \tag{4.3}
\end{equation}
and the binomial weight function 
\begin{equation}
\varrho \left( x\right) =\frac{N!}{\Gamma \left( x+1\right) \Gamma \left(
N-x+1\right) }p^{x}\left( 1-p\right) ^{N-x}.  \tag{4.4}
\end{equation}
The Kravchuk functions can be defined as the ket vectors with wavefunctions 
\begin{equation}
\phi _{k}^{\left( p\right) }\left( x,N\right) :=\left( d_{k}\right) ^{-1}%
\sqrt{\varrho \left( x+Np\right) }K_{k}^{\left( p\right) }\left(
Np+x,N\right)  \tag{4.5}
\end{equation}
where $d_{k}^{2}=\frac{N!}{k!\left( N-k\right) !}\left( p\left( 1-p\right)
\right) ^{k},$ $k\in \left\{ 0,1,...,N\right\} $ and $-Np\leq x\leq \left(
1-p\right) N.$ They obey the following discrete orthogonality relation 
\begin{equation}
\sum_{j=0}^{N}\phi _{k}^{\left( p\right) }\left( x_{j},N\right) \phi
_{n}^{\left( p\right) }\left( x_{j},N\right) =\delta _{n,k}.  \tag{4.6}
\end{equation}
at the points $x_{j}=\left( j-pN\right) .$ Following ($\left[ 8\right] $,
p.370), the functions $\phi _{k}^{\left( p\right) }\left( x,N\right) $ are
eigenfunctions of the Kravchuk oscillator with the Hamiltonian 
\begin{equation}
H^{N}:=2p\left( 1-p\right) N+\frac{1}{2}+\left( 1-2p\right) \frac{\xi }{h}-%
\sqrt{p\left( 1-p\right) }\left( \alpha \left( \xi \right) e^{h\partial
_{\xi }}+\alpha \left( \xi -h\right) e^{-h\partial _{\xi }}\right)  \tag{4.8}
\end{equation}
where 
\begin{equation}
h=\sqrt{2Np\left( 1-p\right) },\alpha \left( \xi \right) =\sqrt{\left(
\left( 1-p\right) N-h^{-1}\xi \right) \left( pN+1+h^{-1}\xi \right) }. 
\tag{4.9}
\end{equation}
This operator is acting in the Hilbert space $l^{2}\left( \xi \right) $ with
orthonormal basis consisting of Kravchuk functions $\left( 4.5\right) $
which verify 
\begin{equation}
H\phi _{k}^{\left( p\right) }\left( x,N\right) =\left( k+\frac{1}{2}\right)
\phi _{k}^{\left( p\right) }\left( x,N\right) ,\text{ }k=0,1,...,N. 
\tag{4.10}
\end{equation}
It have been also pointed out $\left[ 8\right] $ that these functions
coincide with the harmonic oscillator functions in the limit as $%
N\rightarrow \infty ,$ namely 
\begin{equation}
\lim_{N\rightarrow \infty }h^{-\frac{1}{2}}\phi _{k}^{\left( p\right)
}\left( h^{-1}\xi ,N\right) =\left( \sqrt{\pi }2^{k}k!\right) ^{-\frac{1}{2}%
}H_{k}\left( \xi \right) e^{-\frac{1}{2}\xi ^{2}}  \tag{4.11}
\end{equation}
where $H_{k}\left( .\right) $ are the Hermite polynomials; see also $\left( 
\left[ 20\right] ,\text{p.133}\right) .$

Finally, in the subsequent we will use the notation $q=1-p$ and we will be
concerned with the Kravchuk functions 
\begin{equation}
\phi _{k}^{\left( p,q\right) }\left( x,N\right) :=K_{k}^{\left( p\right)
}\left( x+Np,N\right) \sqrt{\frac{k!\left( N-k\right) !p^{Np+x}q^{Nq-x}}{%
p^{k}q^{k}\Gamma \left( Np+x+1\right) \Gamma \left( Nq-x+1\right) }} 
\tag{4.12}
\end{equation}
\textbf{Remark.4.1.} The normalized Kravchuk function in $\left( 4.12\right) 
$ can also be expressed in terms of the Wigner $\mathbf{d}$-function by $%
\left( -1\right) ^{s-r}\mathbf{d}_{s,r}^{j}\left( \beta \right) \equiv \phi
_{k}^{\left( p,q\right) }\left( x,N\right) $ where $j=N/2,$ $k=j-s,x=j-r$
and $p=\sin ^{2}(\beta /2);$ see $\left[ 21\right] -\left[ 22\right] .$ Note
also that the group theoretical interpretation of the dynamical $su(2)$
algebra for the Kravchuk functions can be found in $\left[ 23\right] .$ For
more details we refer to $\left[ 24\right] $ and references therein.

\textbf{Remark 4.2. }For $p=1/2$ it is interesting $\left[ 23\right] $ that
in this situation the double commutator of the Hamiltonian $\left(
4.8\right) $ with the variable $x$ satisfies the equation $\left[
H^{N}\left( x\right) ,\left[ H^{N}\left( x\right) ,x\right] \right] =x$
which can be viewed as the difference analogue of the equation of motion of
the linear harmonic oscillator in the Schr\"{o}dinger representation as
pointed out in ($\left[ 8\right] ,p.371$ ).

\section{Generalized spin coherent states for the Kravchuk oscillator}

We now define a class of generalized spin coherent states (GSCS) for the
Kravchuk oscillator according to definition $\left( 3.1\right) $ as follows.

\textbf{Definition 5.1. }\textit{For }$m\in \mathbb{Z}_{+}$, $2\nu =1,2,...,$
\textit{and} $0<p<1$ \textit{with} $q=1-p$\textit{. A class of GSCS\ for the
Kravchuk oscillator are defined by }

\begin{equation}
\mid z,\nu ,m>_{\left( p,q\right) }:=\left( \mathcal{N}_{\nu ,m}\left(
z\right) \right) ^{-\frac{1}{2}}\sum_{k=0}^{2(\nu +m)}\overline{\widetilde{%
\Phi }_{k}^{(\nu ,m)}\left( z\right) }\mid \phi _{k}^{\left( p,q\right)
}\left( \bullet ,2(\nu +m)\right) >  \tag{5.1}
\end{equation}
\textit{where }$\mathcal{N}_{\nu ,m}\left( z\right) $\textit{\ is the factor
in (3.12)}$,$\textit{\ }$\widetilde{\Phi }_{k}^{(\nu ,m)}\left( z\right) $%
\textit{\ are the monopole harmonics defined in (2.15) and }$\phi
_{k}^{\left( p,q\right) }\left( \bullet ,2(\nu +m)\right) $ \textit{are the
Kravchuk functions (4.12)\ with }$N=2\left( \nu +m\right) .$

\textbf{Proposition 5.1.}\textit{The wavefunction of these GSCS in (5.1) are
of the form} 
\begin{eqnarray}
&<&x\mid z,\nu ,m>_{\left( p,q\right) }=\frac{\left( 2\nu +2m\right) !}{%
\overline{z}^{m}\left( 1+z\overline{z}\right) ^{\nu }}\sqrt{\frac{\left(
2\nu +m\right) !m!p^{2\left( \nu +m\right) p+x}q^{2\left( \nu +m\right) q-x}%
}{\Gamma \left( 2\left( \nu +m\right) p+x+1\right) \Gamma \left( 2\left( \nu
+m\right) q-x+1\right) }}  \TCItag{5.2} \\
&&\times \sum_{k=0}^{2\left( \nu +m\right) }\frac{\left( -1\right) ^{k}%
\overline{z}^{k}P_{k}^{\left( -2\left( \nu +m\right) -1,-x+2\left( \nu
+m\right) q-k\right) }\left( 1-\frac{2}{p}\right) }{\left( -2\left( \nu
+m\right) \right) _{k}(2\left( \nu +m\right) -k)!}\sqrt{\frac{p^{k}}{q^{k}}}%
P_{m}^{\left( k-m,2\nu +m-k\right) }\left( \frac{1-z\overline{z}}{1+z%
\overline{z}}\right)  \notag
\end{eqnarray}

\textbf{Proof. }We start from Eq.(5.1) by writing 
\begin{equation}
<x\mid z,\nu ,m>_{\left( p,q\right) }=\left( N+1\right) ^{-\frac{1}{2}%
}\sum_{k=0}^{N}\overline{\widetilde{\Phi }_{k}^{(\nu ,m)}\left( z\right) }%
\phi _{k}^{\left( p,q\right) }\left( x,N\right)  \tag{5.3}
\end{equation}
where the\ monopole harmonic function 
\begin{equation}
\widetilde{\Phi }_{k}^{(\nu ,m)}\left( z\right) =\sqrt{\frac{\left(
N+1\right) \left( 2\nu +m\right) !m!}{k!(N-k)!}}\frac{z^{k-m}}{\left( 1+z%
\overline{z}\right) ^{\nu }}P_{m}^{\left( k-m,N-m-k\right) }\left( u\right) 
\tag{5.4}
\end{equation}
is obtained from the expression $\widetilde{\Phi }_{j}^{(\nu ,m)}\left(
z\right) $, $-m\leq j\leq 2\nu +m$ by setting $k=m+j$ and $u=\left( 1-z%
\overline{z}\right) \left( 1+z\overline{z}\right) ^{-1}.$ For $k\in \left\{
0,1,2,...,N\right\} $ and $-\frac{N}{2}\leq x\leq \frac{N}{2}$. If the
Kravchuk polynomial is expressed in terms of the $_{2}\digamma _{1}$-sum by
using $\left( 4.2\right) ,$then we can rewrite the functions $\phi
_{k}^{\left( p,q\right) }\left( .\right) $ as 
\begin{eqnarray}
\phi _{k}^{\left( p,q\right) }\left( x,N\right) &=&\frac{\left( -1\right)
^{k}N!}{\sqrt{k!\left( N-k\right) !}}\sqrt{\frac{p^{k}q^{-k}p^{Np+x}q^{Nq-x}%
}{\Gamma \left( Np+x+1\right) \Gamma \left( Nq-x+1\right) }}  \TCItag{5.5} \\
&&._{2}\digamma _{1}\left( -k,-\left( x+Np\right) ,-N;\frac{1}{p}\right) . 
\notag
\end{eqnarray}
We make appeal to the relation $\left( 2.13\right) $ in order to write the $%
_{2}\digamma _{1}-$sum in terms of the Jacobi polynomial as

\begin{equation}
_{2}\digamma _{1}\left( -k,-\left( x+Np\right) ,-N;\frac{1}{p}\right) =\frac{%
k!}{\left( -N\right) _{k}}P_{k}^{\left( -N-1,-x-k+Nq\right) }\left( 1-\frac{2%
}{p}\right) .  \tag{5.6}
\end{equation}
Therefor, Eq.$\left( 5.5\right) $ takes the form 
\begin{eqnarray}
\phi _{k}^{\left( p,q\right) }\left( x,N\right) &=&\frac{\left( -1\right)
^{k}N!}{\sqrt{\left( N-k\right) !}}\sqrt{\frac{p^{k}q^{-k}k!p^{Np+x}q^{Nq-x}%
}{\Gamma \left( Np+x+1\right) \Gamma \left( Nq-x+1\right) }}  \TCItag{5.7} \\
&&\times \frac{1}{\left( -N\right) _{k}}P_{k}^{\left( -N-1,-x-k+Nq\right)
}\left( 1-\frac{2}{p}\right) .  \notag
\end{eqnarray}
Returning back to $\left( 5.3\right) $, we get successively 
\begin{equation}
\frac{1}{\sqrt{N+1}}\sum_{k=0}^{N}\overline{\widetilde{\Phi }_{k}^{(\nu
,m)}\left( z\right) }\phi _{k}\left( x,N\right) =\sum_{k=0}^{N}\sqrt{\frac{%
\left( 2\nu +m\right) !m!}{k!(N-k)!}}\frac{\overline{z}^{k-m}}{\left( 1+z%
\overline{z}\right) ^{\nu }}P_{m}^{\left( k-m,N-m-k\right) }\left( u\right) 
\tag{5.8}
\end{equation}
\begin{equation*}
\times \frac{\left( -1\right) ^{k}N!}{\sqrt{\left( N-k\right) !}}\sqrt{\frac{%
k!p^{Np+x+k}q^{Nq-x-k}}{\Gamma \left( Np+x+1\right) \Gamma \left(
Nq-x+1\right) }}\frac{1}{\left( -N\right) _{k}}P_{k}^{\left(
-N-1,-x-k+Nq\right) }\left( 1-\frac{2}{p}\right)
\end{equation*}
\begin{equation}
=\frac{N!\overline{z}^{-m}}{\left( 1+z\overline{z}\right) ^{\nu }}\sqrt{%
\frac{\left( 2\nu +m\right) !m!p^{Np+x}\left( 1-p\right) ^{Nq-x}}{\Gamma
\left( Np+x+1\right) \Gamma \left( Nq-x+1\right) }}  \tag{5.9}
\end{equation}
\begin{equation*}
\times \sum_{k=0}^{N}\frac{\left( -1\right) ^{k}\overline{z}^{k}}{\left(
-N\right) _{k}(N-k)!}\sqrt{\frac{p^{k}}{q^{k}}}P_{m}^{\left(
k-m,N-m-k\right) }\left( u\right) P_{k}^{\left( -N-1,-x+Nq-k\right) }\left(
1-\frac{2}{p}\right) .
\end{equation*}
By Eq.$\left( 5.9\right) $ we arrive at the expression $\left( 5.2\right) .$ 
$\blacksquare $

Now, keeping $N=2\left( \nu +m\right) $ and denoting by $\mathcal{H}$ the ($%
2N+1)-$dimensional Hilbert space generated by the Kravchuk eigenstates then
we can construct ''\textit{\`{a} la Bargmann}'' $\left[ 25\right] $ for any
state vector $\mid \phi >$ in $\mathcal{H}$ the corresponding function in
the eigenspace $\mathcal{A}_{m}^{\nu }\left( \mathbb{S}^{2}\right) $ defined
in $\left( 2.12\right) .$ This function is the Bargmann transform of the
state $\mid \phi >$ , say $\mathcal{B}_{\nu ,m}\left[ \phi \right] $ , which
is performed by applying the coherent state transform formalism $\left[ 4%
\right] $. Precisely, for each $m\in \mathbb{Z}_{+}$ , it is defined as $%
\mathcal{B}_{\nu ,m}:\mathcal{H}\rightarrow \mathcal{A}_{m}^{\nu }\left( 
\mathbb{S}^{2}\right) $ by 
\begin{equation}
\mathcal{B}_{\nu ,m}\left[ \phi \right] \left( z\right) :=\left( 2(\nu
+m)+1\right) ^{\frac{1}{2}}\left\langle \phi \mid z,\nu ,m\right\rangle _{%
\mathcal{H}}.  \tag{5.10}
\end{equation}
Thus, Eq.$\left( 5.10\right) $ together with Eq.$\left( 3.16\right) $ lead
to the following representation of any state vector $\mid \phi >$ in $%
\mathcal{H}$ in terms of the constructed GSCS $\mid z,\nu ,m>$ as 
\begin{equation}
\mid \phi >=\int_{\mathbb{C}}d\mu _{\nu ,m}\left( z\right) \left( 2(\nu
+m)+1\right) ^{-\frac{1}{2}}\mathcal{B}_{\nu ,m}\left[ \phi \right] \left(
z\right) \mid z,\nu ,m>.  \tag{5.11}
\end{equation}
Finally, taking into account Eq.$\left( 3.17\right) ,$ we obtain from $%
\left( 5.11\right) $ the equality 
\begin{equation}
\left\langle \phi \mid \phi \right\rangle _{\mathcal{H}}=\int_{\mathbb{C}%
}\left| \mathcal{B}_{\nu ,m}\left[ \phi \right] \left( z\right) \right|
^{2}G_{11}^{11}\left( z\overline{z}\mid 
\begin{array}{c}
-1 \\ 
0
\end{array}
\right) d\eta \left( z\right) ,  \tag{5.12}
\end{equation}
where the Meijer's G-function and the Lebesgue measure $d\eta \left(
z\right) $ are employed. These notation could be of help when tackling the
photon-added coherent states problem for the constructed GSCS $\mid z,\nu
,m> $ as mentioned in a previous remark.

\textbf{Remark 3.1.}We should note that a set of coherent states attached to
spherical Landau levels, which form is similar to $\left( 5.1\right) ,$ have
been performed in $\left[ 6\right] $ and $\left[ 11\right] $ $-\left[ 26%
\right] $ with the choice for the Hilbert space carrying them as the space
of polynomials of degree less than $2\nu +2m+1$ endowed with an orthonormal
basis of the form: $\phi _{j}\left( \varkappa \right) =\left( \left( 2\nu
+2m!\right) \left( \left( 2\nu +m-j\right) !\right) ^{-1}\left( \left(
j+m\right) !\right) ^{-1}\right) ^{\frac{1}{2}}\varkappa ^{j+m}$ where $%
\varkappa \in \mathbb{C}$ and $0\leq j\leq 2\left( \nu +m\right) .$

\section{The case m=0}

Now, from the above proposition 5.1 , we can deduce the following result.

\textbf{Corollary 6.1.}\textit{\ For}\textbf{\ }$2\nu =1,2,...,$ \textit{and}
$0<p<1$ \textit{with} $q=1-p$\textit{.}\textbf{\ }\textit{The wave functions
for GSCS in (5.2), corresponding to the lowest spherical Landau level }$%
\lambda _{\nu ,0}$ in \textit{(2.5)\ are of the form}

\begin{eqnarray}
&<&x\mid z,\nu ,0>_{\left( p,q\right) }=\frac{\sqrt{N!}}{\left( 1+z\overline{%
z}\right) ^{\nu }}\sqrt{\frac{p^{Np+x}q^{Nq-x}}{\Gamma \left( Np+x+1\right)
\Gamma \left( Nq-x+1\right) }}  \TCItag{6.1} \\
&&\times \left( 1+\sqrt{\frac{q}{p}}\overline{z}\right) ^{x+Np}\left( 1-%
\sqrt{\frac{p}{q}}\overline{z}\right) ^{Nq-x}  \notag
\end{eqnarray}
\textit{where} $N=2\nu .$

\textbf{Proof. }We start by putting $m=0$ in the expression $\left(
5.3\right) .$ That is 
\begin{equation}
<x\mid z,\nu ,0>_{\left( p,q\right) }=\left( N+1\right) ^{-\frac{1}{2}%
}\sum_{k=0}^{N}\overline{\widetilde{\Phi }_{k}^{(\nu ,0)}\left( z\right) }%
\phi _{k}^{\left( p,q\right) }\left( x,N\right) .  \tag{6.2}
\end{equation}
Using the fact that $P_{0}^{\left( \alpha ,\beta \right) }\left( u\right) =1$%
, then Eq.$\left( 5.4\right) $ reduces to 
\begin{equation}
\widetilde{\Phi }_{k}^{(\nu ,0)}\left( z\right) =\sqrt{\frac{\left(
N+1\right) N!}{k!(N-k)!}}\frac{z^{k}}{\left( 1+z\overline{z}\right) ^{\nu }}.
\tag{6.3}
\end{equation}
Replacing the expression $\left( 6.3\right) $ in Eq.$\left( 6.2\right) ,$ we
obtain that 
\begin{equation}
<x\mid z,\nu ,0>_{\left( p,q\right) }=\frac{\sqrt{N!}}{\left( 1+z\overline{z}%
\right) ^{\nu }}\sum_{k=0}^{N}\sqrt{\frac{1}{k!(N-k)!}}\overline{z}^{k}\phi
_{k}^{\left( p,q\right) }\left( x,N\right) .  \tag{6.4}
\end{equation}
On the other hand, we make use of Eq.$\left( 5.5\right) ,$ to rewrite $%
\left( 6.4\right) $ as 
\begin{equation}
<x\mid z,\nu ,0>_{\left( p,q\right) }=\frac{\sqrt{N!}}{\left( 1+z\overline{z}%
\right) ^{\nu }}\sqrt{\frac{p^{Np+x}q^{Nq-x}}{\Gamma \left( Np+x+1\right)
\Gamma \left( Nq-x+1\right) }}  \tag{6.5}
\end{equation}
\begin{equation*}
\times \sum_{k=0}^{N}\left( -\overline{z}\sqrt{\frac{p}{q}}\right)
^{k}.\left( 
\begin{array}{c}
N \\ 
k
\end{array}
\right) ._{2}\digamma _{1}\left( -k,-\left( x+Np\right) ,-N;\frac{1}{p}%
\right) .
\end{equation*}
Now, with the help of the generating function $\left( \left[ 27\right]
,p.184\right) :$%
\begin{equation}
\left( 1-\frac{q}{p}t\right) ^{\zeta }\left( 1+t\right) ^{N-\zeta
}=\sum\limits_{n=0}^{N}\left( 
\begin{array}{c}
N \\ 
k
\end{array}
\right) K_{k}\left( \zeta ,p,N\right) t^{n}.  \tag{6.6}
\end{equation}
where $\zeta =0,1,2,...,N$ and $K_{k}\left( x,p,N\right) =_{2}F_{1}\left(
-k,-x,-N;\frac{1}{p}\right) $. We apply it for $t=$ $-\overline{z}\sqrt{%
\frac{p}{q}}$ and $\zeta =x+Np$ to find that 
\begin{eqnarray}
&&\sum_{k=0}^{N}\left( -\overline{z}\sqrt{\frac{p}{q}}\right) ^{k}\left( 
\begin{array}{c}
N \\ 
k
\end{array}
\right) ._{2}\digamma _{1}\left( -k,-\left( x+Np\right) ,-N;\frac{1}{p}%
\right)  \TCItag{6.7} \\
&=&\left( 1+\sqrt{\frac{q}{p}}\overline{z}\right) ^{x+Np}\left( 1-\sqrt{%
\frac{p}{q}}\overline{z}\right) ^{Nq-x}.  \notag
\end{eqnarray}
Finally, in view of $\left( 6.7\right) $, Eq.$\left( 6.5\right) $ takes the
form 
\begin{equation}
<x\mid z,\nu ,0>_{\left( p,q\right) }=\frac{\sqrt{N!}}{\left( 1+z\overline{z}%
\right) ^{\nu }}\sqrt{\frac{p^{Np+x}q^{Nq-x}}{\Gamma \left( Np+x+1\right)
\Gamma \left( Nq-x+1\right) }}\left( 1+\sqrt{\frac{q}{p}}\overline{z}\right)
^{Np+x}\left( 1-\sqrt{\frac{p}{q}}\overline{z}\right) ^{Nq-x}  \tag{6.8}
\end{equation}
\textbf{\ } This ends the proof of the corollary.$\blacksquare $

We should note that in $\left[ 9\right] $ Chenaghlou and Faizy have
constructed a class of Klauder-Perelomov coherent states by acting on the
ground state function 
\begin{equation}
\psi _{0}\left( y\right) =\sqrt{\frac{N!p^{y}\left( 1-p\right) ^{N-y}}{%
y!\left( N-y\right) !}},0<p<1  \tag{6.9}
\end{equation}
via a displacement operator defined by two generators of the Lie algebra $%
so\left( 3\right) .$ The wave functions of their coherent states is were of
the form$\left( \left[ 9\right] ,\text{ Eq.(38)}\right) $: 
\begin{equation}
<y\mid z,N>^{KP}:=\left( 1+\frac{p}{1-p}z\overline{z}\right) ^{-\frac{1}{2}%
N}\left( 1+z\right) ^{y}\left( 1-\frac{pz}{1-p}\right) ^{N-y}\sqrt{\frac{%
N!p^{y}\left( 1-p\right) ^{N-y}}{y!\left( N-y\right) !}}.  \tag{6.10}
\end{equation}
To establish a connection between the coherent states $\mid z,N>^{KP}$in $%
\left( 6.10\right) $ \ and our constructed coherent states in the case $m=0$
we need to make a little change of variables in Eq.$\left( 6.1\right) .$ We
precisely consider the following replacements: $x\rightarrow y-Np$ and $%
z\rightarrow \sqrt{\frac{p}{q}}\overline{z}$.$\ $By this way, one can easily
check that the following fact: 
\begin{equation}
<y-Np\mid \sqrt{\frac{p}{q}}\overline{z},\nu ,0>_{\left( p,q\right) }=<y\mid
z,N>^{KP}.  \tag{6.11}
\end{equation}
Finally, if one particularize the unitary transform $\left( 5.10\right) $
for the case $m=0$ then one can recover the \textit{analytic} coherent
states representation of the any state vector $\mid \phi >$ in $\mathcal{H}$
as discussed in $\left[ 9\right] .$

\begin{center}
\textbf{References}
\end{center}

\begin{quote}
$\left[ 1\right] ${\small \ J. M. Radcliffe, Some properties of coherent
spin states, \textit{J. Phys. A: Gen Phys.} \textbf{4} (1971) pp.313-323}

$\left[ 2\right] ${\small \ R. Gilmore, Geometry of symmetrized states, 
\textit{Ann. Phys. (NY)}, \textbf{74}, p.391 (1972)}

$\left[ 3\right] ${\small \ A. Perelomov, Coherent States for Arbitrary Lie
group, \textit{Commun. Math. Phys.}\textbf{\ 26}, pp.222-236 (1972)\newline
}

$\left[ 4\right] ${\small \ J. P. Gazeau, Coherent states in quantum
physics, WILEY-VCH Verlag GmbH \& Co. KGaA Weinheim 2009}

$\left[ 5\right] ${\small \ W. Feller, An introduction to probability:
theory and its applications, Vol1 2nd ed., John Wiley, 1957}

$\left[ 6\right] ${\small \ A. Ghanmi, A. Hafoud and Z. Mouayn, Generalized
binomial probability distributions attached to Landau levels on the Riemann
sphere,\textit{\ Adv. Math. Phys.} Vol 2011, Article ID393417}

$\left[ 7\right] ${\small \ J. R. Klauder, Continuous Representation theory
I. Postulates of continuous representation theory, \textit{J. Math. Phys}. 
\textbf{4 }\ pp.1055-1058}

$\left[ 8\right] ${\small \ N. M. Atakishieyev and K. B. Wolf, Approximation
on a finite set of points through Kravchuk functions, \textit{Revista
Mexicana de F\'{i}sica}.\ \textbf{40}, No. 3 pp.366-377 (1994)}

$\left[ 9\right] ${\small \ A. Chenaghlou and O. Faizy, Barut-Girardello and
Klauder-Perelomov coherent states for the Kravchuk functions, \textit{J.
Math. Phys.} \textbf{48}, 112106 (2007)}

$\left[ 10\right] ${\small \ E. V. Ferapontov and A.P. Veselov: Integrable
Schr\"{o}dinger operators with magnetic fields: Factorization method on
curved surfaces, \textit{J. Math. Phys.} \textbf{42} (2001), 590-607.}

$\left[ 11\right] ${\small \ Z. Mouayn, Coherent states attached to Landau
levels on the Riemann sphere, \textit{Rep. Math. Phys.}\textbf{\ 55},
pp.269-276 No.2 (2005)}

$\left[ 12\right] ${\small \ T. T. Wu TT and C. N. Yang , Dirac monopole
without strings: monopole harmonics, \textit{Nucl. Phys B}. \textbf{107},
pp.364-380 (1976)}

$\left[ 13\right] ${\small \ J. Peetre and G. Zhang, Harmonic analysis on
the quantized Riemann sphere,\textit{\ Internat. J. Math. \& Math. Sci }\ 
\textbf{16}, No 2, 225-244 (1993)}

$\left[ 14\right] ${\small \ G. Szeg\"{o}, Orthogonal polynomials. American
Mathematical Society; Providence, R.I. (1975)\newline
}

$\left[ 15\right] ${\small \ F. Hirzebruch, ''Topological Methods in
Algebraic Geometry'', 131, Grundlehren der mathematischen Wissenschaften,
Sringer-Verlag, 1978}

$\left[ 16\right] ${\small \ D. J. Simms and N.M. Woodhouse, lectures on
Geometric Quantization, Lectures Notes in Physics, Vol.53, Springer-Verlag,
Berlin 1976}

$\left[ 17\right] ${\small \ Y. M. Shnir, Magnetic Monopoles, texts and
monographs in Physics, Springer-Verlag Berlin Heidelberg 2005}

$\left[ 18\right] ${\small \ D. Stoler, B.E.A. Saleh and M.C. Teich,
Binomial states of the quantized radiation field. \textit{Optic. Acta}.%
\textbf{\ 32}, no.3, pp.345-355 (1985)}

$\left[ 19\right] ${\small \ A. M. Mathai and R. K. Saxena, Generalized
hypergeometric functions with applications in statistics and physical
sciences, Lect. Notes. Math. Vol 348, Springer-Verlag, Berlin, 1973}

$\left[ 20\right] ${\small \ A. Nikiforov and V. Ouvarov, Fonctions
speciales de la physique math\'{e}matique, Edition Mir (1983)\newline
}

$\left[ 21\right] ${\small \ C. Campigotto, Yu. F. Smirnov and S. G.
Enikeev, }$q${\small -Analog of the Kravchuk and Meixner orthogonal
polynomials\textit{. J. Comp. Appl. Math.} \textbf{57} pp.87-97 (1995)}

$\left[ 22\right] ${\small \ A. F. Nikiforov, S. K. Suslov and V. B.
Ouvarov, Classical orthogonal polynomials of a discrete variable. Moscow,
Nauka (1985)}

$\left[ 23\right] ${\small \ N. M. Atakishieyev and S. K. Suslov, Difference
analogs of the harmonic oscillator. \textit{Theor. Math. Phys.} \textbf{85} }%
$\left( 1991\right) ${\small \ p.1055}

$\left[ 24\right] ${\small \ Yu F. Smirnov, On factorization and
algebraization of difference equations of hypergeometric type, M. Alfaro et
al. (Eds) Proceeding of the International Workshop on Orthogonal Polynomials
in Mathematical Physics. Legan\'{e}s, 24-26 june, 1996. M. Alfaro et al.
(Eds) (1997)}

$\left[ 25\right] ${\small \ V. Bargmann, On a Hilbert space of analytic
functions and an associated integral transform, \textit{Part I. Comm. Pure.
Appl. Math}., \textbf{14 }187-214 (1961)}

$\left[ 26\right] ${\small \ Z. Mouayn, Coherent states attached to the
spectrum of the Bochner Laplacian for the Hopf fibration. \textit{J. Geo \&
Phys}. \textbf{59}, \ Issue 2, pp.256-261 (2009)}

$\left[ 27\right] ${\small \ Mourad E.H.Ismail, Classical and Quantum
Orthogonal Polynomials in one variable, Encyclopedia of Mathematics and its
applications, Cambridge university press (2005)}
\end{quote}

\end{document}